\documentclass{article}
\usepackage{emulateapj,graphics,onecolfloat}

\font\tfa=cmr10 at 8.00pt
\tfa

\begin{document}
\twocolumn[
\title {Studying clusters with Planck}
\author {Martin White}
\affil{Departments of Physics and Astronomy, University of California,
Berkeley, CA 94720}

\begin{abstract}
\noindent
\rightskip=0pt
We use mock Sunyaev-Zel'dovich effect (SZE) maps to investigate how well
the {\sl Planck\/} mission might find and characterize clusters of galaxies.
We discuss different combinations of frequency maps and different methods
for identifying cluster candidates.  For the simplest methods, the catalogues
are not complete even for relatively high mass thresholds, but the full sky
nature of the mission ensures a large sample of massive, high-$z$ clusters
which will be ideal for many studies.  We make a preliminary attempt to
identify the X-ray, optical and weak lensing properties of the {\sl Planck\/}
sample.
\end{abstract}

]

\rightskip=0pt

\section{Introduction}

The {\sl Planck\/} mission\footnote{http://astro.estec.esa.nl/Planck/} will
provide all-sky maps of superb resolution at 9 frequencies, ranging
{}from $30\,$GHz to $850\,$GHz, with unprecedented signal-to-noise.
One of the science goals that this enables is the construction of a catalogue
of galaxy clusters detected through the Sunyaev-Zel'dovich effect
(SZE: Sunyaev \& Zel'dovich~(\cite{SZ72,SZ80}); for recent reviews see
Rephaeli~\cite{Rep} and Birkinshaw~\cite{Bir})
and a study of the residual SZ emission once these clusters are removed.
In this paper we investigate some of the ways a survey with the
characteristics of the {\sl Planck\/} survey will advance our understanding
of cluster physics and cosmology, building on the earlier work of many authors
(Aghanim et al.~\cite{Aghanim}; Kay, Liddle \& Thomas \cite{KayLidTho};
 Vielva et al.~\cite{VBHMLST}; Herranz et al.~\cite{HSHBDML};
 Diego et al.~\cite{Diego}; Hobson \& McLachlan \cite{HobMcL}).
We discuss the sample {\sl Planck\/} will select,
what observations it can give of previously known clusters, and some of
the properties of the clusters which will be useful for followup.

Exploring the expected {\sl Planck\/} sample is a complex problem, involving
several different but interlocking facets.  Our aim here is to focus on one,
while treating the others in as simple a manner as is consistent with what we
know.  This investigation is thus by nature preliminary, but we believe
that the simulations upon which it is based provide an advance over what has
been used in the past and illuminate several issues future {\sl Planck\/} work
will need to address.  We use mock SZ maps drawn from a large volume, high
resolution N-body simulation.  These maps capture much of the physics behind
the SZ effect, with the sources (groups and clusters) situated correctly in
their cosmological context.  Due to the current uncertainty in the amplitude
of both the SZ signal and the relevant astrophysical foregrounds, a detailed
modeling of the `noise' is not possible.  However we shall investigate
signal-to-noise levels which should span the allowed range.
Our investigations lead us to conclude that constructing a catalogue of
clusters from {\sl Planck\/} will be a difficult undertaking, but one which
is worth a great deal of effort.  Which are the optimal methods for
foreground subtraction, cluster finding and modeling the selection function
are all open questions currently, and this work will only begin to explore
some of the relevant issues.  Also uncertain is the most efficient way of
collecting relevant information during the pre-launch phase of
{\sl Planck}.

The outline of the paper is as follows.  In \S\ref{sec:hifreq} we describe
our methodology for isolating the SZ signal from the other astrophysical
components (including the CMB) in our simulated observations, and the
simulations we are using to model the SZ effect itself.  We describe two
methods for finding cluster candidates in our simulated maps in
\S\ref{sec:find} and how well they do in \S\ref{sec:results}.
We discuss using the lower frequency channels of {\sl Planck\/} in order
to reduce IR point source contamination in \S\ref{sec:lowfreq}.
In \S\ref{sec:followup} we describe the properties of our cluster sample
{}from the point of view of X-ray, optical and weak lensing followup.
The dependence of our results on our cosmological and foreground modeling
is discussed in \S\ref{sec:model} and our conclusions are presented in
\S\ref{sec:conclusions}.

\section{Simulated observations} \label{sec:hifreq}

\begin{figure}
\begin{center}
\resizebox{3.5in}{!}{\includegraphics{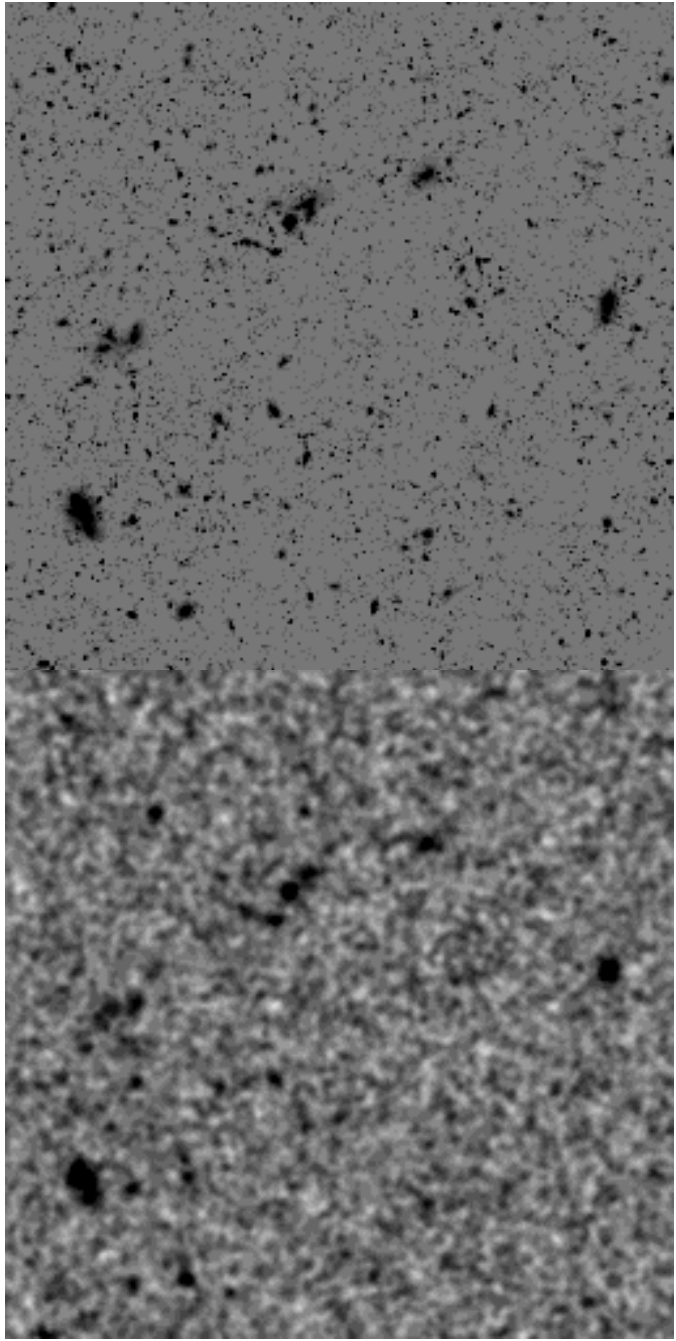}}
\end{center}
\caption{One of our simulated maps: $\Delta T(353)-\Delta T(217)$.
(Top) the input SZ signal, converted to $\Delta T$ at $353\,$GHz but without
noise or smoothing.  (Bottom) the same map, with $40\,\mu$K of noise
and $5'$ smoothing.  In both panels the greyscale is linear in $\Delta T$
ranging from $-100\,\mu$K (white) to $+100\,\mu$K (black).}
\label{fig:maps}
\end{figure}

The thermal SZE signal is likely to be dominated by virialized objects
whose typical angular size is around $1'$.  Thus if we can achieve a
suitably low noise level, including foreground subtraction, the {\sl Planck\/}
channels which will be best suited to studying the SZE signal from galaxy
clusters will be those at higher frequency, where the angular resolution is
best.  If it turns out that high frequency foregrounds are particularly
troublesome, and cannot be removed to a suitable level, we may need to
consider moving to lower frequencies, and resolutions, where the `noise' is
lower.
We shall consider both approaches here, beginning with the high frequency
option.  We shall return to the lower frequency option in \S\ref{sec:lowfreq}.

The first step in studying the SZE is to use the multi-frequency capability
of {\sl Planck\/} to isolate the SZ signal from that of the foregrounds and
the background (see Herranz et al.~\cite{HSHBDML} for a summary of recent
work and Gomez et al.~\cite{A3266} for a recent example using real data).
We expect the foregrounds at high and low frequency to be different, with
only the CMB in common across the channels.
We shall phrase the effects of foreground removal in terms of an increase in
the effective ``noise'' of our SZE maps.
This is appropriate in the sense that finding clusters in {\sl Planck\/} maps
is a very local procedure.  The clusters are (at most) a few arcminutes on a
side and the map covers 40,000 square degrees so mostly what matters is the
S/N in the map at the position of the cluster.
By quantifying the required S/N in this way we are attempting to decouple
cluster finding from the details of foreground subtraction.
Isolating the different parts of the problem in this way is appropriate during
this exploratory stage and our current knowledge of the relevant foregrounds
is sufficiently imprecise that our simplistic treatment offers several
advantages.

Let us first consider the SZE `signal', then the CMB `background' and then
the other astrophysical `foregrounds'.
While in general we expect the foregrounds to have unknown, spatially
varying spectral characteristics, for SZE studies we are aided by the
fact that the spectral shape of our primary CMB `background' and SZE
`signal' are very well known.
In thermodynamic units the thermal SZE scales, for non-relativistic $T_e$, as
\begin{eqnarray}
  {\Delta T\over T} &=&
  \phantom{-2}y \left( x{{\rm e}^x+1\over {\rm e}^x-1}-4 \right) \\
  &\simeq& -2y\qquad \mbox{for }\ x\ll 1\, ,
\end{eqnarray}
where $x=h\nu/kT_{\rm CMB}\simeq \nu/56.85\,$GHz is the dimensionless
frequency, and the second expression is valid in the Rayleigh-Jeans
limit.
The quantity $y$ is known as the Comptonization parameter and is given by
\begin{equation}
  y\equiv \sigma_T\int d\ell\ {n_e k(T_e-T_{\rm CMB})\over m_e c^2}\, ,
\label{eqn:ydef}
\end{equation}
where the integral is performed along the photon path.
Since $T_e\gg T_{\rm CMB}$ the integrand is proportional to the integrated
electron pressure along the line of sight.
For frequencies below approximately $217\,$GHz the SZE is a decrement in the
CMB while above this frequency it is an increment.

Obviously the CMB has a constant temperature in thermodynamic units.
In removing this background we could in principle use many of the
{\sl Planck\/} channels, however the differing beam sizes represents
a potential problem if we wish to work at high angular resolution.
All of the channels above $217\,$GHz have a $5'$ FWHM beam, the lower
frequency channels have beams ranging from $33'$ (30GHz) to $7'$ (143GHz).
The fluctuations measured at any point on the sky by a beam of width
$\theta$ differs from that measured with $5'$ resolution by a `noise'
\begin{equation}
  \Delta T^2 = {1\over 4\pi}\sum_\ell (2\ell+1)C_\ell
    \left( \Delta B_\ell \right)^2
\end{equation}
where $C_\ell$ is the angular power spectrum of the CMB at multipole
moment $\ell$ and $B_\ell$ is the (Legendre) transform of the beam.
For a {\sl COBE\/} normalized $\Lambda$CDM model this typically ranges from 
$5\,\mu$K at $7'$ through $14\,\mu$K at $10'$ to $50\,\mu$K at $33'$
(we use thermodynamic temperatures throughout).
The difference in the CMB signals measured by the $217\,$GHz channel and
frequency channels below $100\,$GHz is thus comparable to the noise
($13\,\mu$K) in the $217\,$GHz channel itself.
Thus only the $143\,$GHz channel provides any strong additional estimate
of the CMB temperature at $5'$ scales, the lower frequency channels serve
primarily to guard against low frequency foregrounds.

At $5'$ {\sl Planck\/} has 4 channels: $217\,$GHz, $353\,$GHz, $545\,$GHz
and $857\,$GHz.
For simplicity we shall neglect the finite width of the {\sl Planck\/}
bands.  If the frequency dependence of the signals is known and the bands
are well characterized (Church, Knox \& White~\cite{ChuKnoWhi}) the bandwidth
presents no additional problems in principle, and we lose nothing by
neglecting it here.

Finally we consider the astrophysical foregrounds, which we assume will be
dominated by extra-galactic point sources at the relevant frequencies.
We therefore implicitly assume that any dust emission has been removed
using a combination of the high frequency channels and/or spatial filtering,
or we are working in a (rare) low dust region of the sky.
If the frequency dependence of the dust is known, or we need only fit a single
extra parameter (e.g.~a power-law index for $h\nu\ll kT_{\rm dust}$) the high
frequency channels form an excellent dust monitor even on a pixel-by-pixel
basis.  In fact, detailed simulations suggest that dust can be removed quite
well from the {\sl Planck\/} maps
(Vielva et al.~\cite{VBHMLST}, Table 2 or
 Stolyarov et al.~\cite{SHAL} Table 2).

The extra-galactic source counts at $353\,$GHz can be taken from observations
with the Submillimeter Common-User Bolometer Array
(SCUBA; Holland et al.~\cite{Holland})
on the James Clerk Maxwell Telescope.
SCUBA has been used to make several deep observations
(Barger et al.~\cite{Barger}, Eales et al.~\cite{Eales},
Holland et al.~\cite{Holletal},
Hughes et al.~\cite{Hughes}, Smail et al.~\cite{SIB}) from which we can
extract source counts.
Using the model in Scott \& White (\cite{ScoWhi}) we predict that
residual point sources should contribute around $70\,\mu$K per
$5'$ beam at $353\,$GHz if we subtract all sources brighter than 4-$5\sigma$
at $545$ and $857\,$GHz.  The source counts, and hence this normalization, are
uncertain at the 50\% level.  An additional uncertainty comes from the fact
that the sources relevant to {\sl Planck\/} are the bright end of the SCUBA
population.  Since SCUBA has surveyed only a small area of sky, our
extrapolations are uncertain, especially if the nature of the brightest
sources differs from the majority of SCUBA sources.

We implicitly assume that the sources are uncorrelated with the clusters
that {\sl Planck\/} will detect.  At low frequencies we know radio point
sources can be correlated with clusters, but we do not know much about
the sub-population which could be detected above $100\,$GHz.
Extrapolations from the recent {\sl WMAP\/} data (Bennett et al.~\cite{WMAP})
suggest radio sources will be a negligible contaminant above $100\,$GHz.
If the IR sources arise from high redshift, as current theories suggest, they
are likely to be uncorrelated with massive clusters.
However this assumption is an idealization which needs to be revisited in
future work.

Our procedure is as follows: of these 4 high resolution bands we will use
the two higher frequency maps to veto or remove the foregrounds and
brightest point sources (if their spectrum is known).
We use the first two bands to isolate the CMB signal from the SZE signal.
For primary CMB anisotropies and SZ signal, the signal-independent,
unbiased, minimum variance estimate of the SZ component is simply the
difference\footnote{If the channels have finite width a recalibration is also
required, but this is a technical detail which need not concern us here.} of
the two channels, as we would have naively guessed.
We will therefore use the simplest possible filtering scheme: differencing
the $353\,$GHz and $217\,$GHz channels.
The difference map is dominated by the noise in the $353\,$GHz channel,
which is roughly $40\,\mu$K per beam.  We assume that foreground
contamination in the $217\,$GHz channel is dominated by that in the $353\,$GHz
channel and that the foregrounds become worse as we go to higher frequencies.

Specifically our first results will be drawn from maps which consist of SZE,
converted to $\Delta T$ at $353\,$GHz and smoothed to an angular resolution
of $5'$, plus noise.
Consistent with our philosophy of isolating different physical effects,
we model the pixel noise as Gaussian, white noise and choose three
noise\footnote{As we discuss in \S\ref{sec:model} our uncertainties in
both signal and noise argue against a more refined set of values at this
point.} levels: $20\,\mu$K, $40\,\mu$K and $80\,\mu$K.
The first corresponds to the noise near the poles where the integration time
is the largest and we assume the sky is abnormally clean of point sources or
perhaps if we additionally make use of the $143\,$GHz channel.
It is likely a lower limit to the noise once additional complicating effects
are included in the analysis.
The second is our fiducial noise from the difference of the $353\,$GHz and
$217\,$GHz channels.  The third includes the contribution from extra-galactic
point sources, using the model above, assuming that the sources which are
not detected at $545$ or $857\,$GHz can be modeled simply as additional
Gaussian noise at $353\,$GHz.  The assumption of Gaussianity implicitly
assumes a `central limit' argument, the treatment of the additional signal
as noise assumes the foregrounds are uncorrelated with the CMB or SZE signal
and that we don't know the true sky monopole (and so subtract a mean from all
of the maps).
Our treatment should therefore be regarded as a simple 1-parameter
``effective'' noise level which will set a target for foreground removal
methods to meet.
We do not consider higher levels of noise/foregrounds as the cosmological
signal becomes progressively harder to extract.

We construct maps of the SZE effect at various frequencies using the method
outlined in Schulz \& White (\cite{SchWhi}).
The maps are created from a large volume, high resolution N-body simulation
containing a `fair sample' of the universe, in a manner tuned to reproduce
the results of the hydrodynamic simulations reported in
White, Hernquist \& Springel (\cite{WHS}).
The normalization has been adjusted to pass through the lower envelope of
the CBI deep field results reported in Mason et al.~(\cite{CBI}) and through
the BIMA point (Dawson et al.~\cite{BIMA}) at higher $\ell$.
This is close to the power seen in more recent CBI data
(A.~Readhead, private communication) and a factor of approximately 4 (in power)
larger than would be predicted by the simulations of the `concordance'
cosmology reported in White, Hernquist \& Springel (\cite{WHS}) -- we shall
discuss lower normalizations in \S\ref{sec:model}.
The N-body maps are not as realistic as those produced using full hydrodynamic
simulations, but they come from a larger volume to provide a better sample of
the rare, high mass clusters.
By using such simulations we ensure that our SZE sources are situated in their
proper cosmological context, including their relation to the overall growth of
large-scale structure.
We shall return to this issue in \S\ref{sec:model}.

We generate 10 maps, each $10^\circ\times 10^\circ$ with $1024^2$ pixels,
and for each map we produce from the underlying simulation a catalogue of
the 3D positions and properties of the clusters in the field.  While the
maps, being drawn from the same underlying simulation, are not fully
independent, they sample different line-of-sight projections and cluster
orientations.  This is important in that our model contains aspherical
clusters with a range of sizes, sub-structures and properties in a range
of environments, providing a more realistic model of what {\sl Planck\/}
should see than randomly placed, sphericized, analytic models.

To the maps we add Gaussian noise scaled to the pixel scale from the levels
quoted above.  Our noise is uniform and white, which will not be the case
for the {\sl Planck\/} mission.  However the noise variations are expected
to be on larger scales than are relevant for finding point-sources, so our
approximation shouldn't affect our results.
An example of one of our maps, additionally smoothed by the beam to reduce
the noise level per $\ll 5'$ pixel, is shown in Fig.~\ref{fig:maps}.

\section{Finding clusters} \label{sec:find}

We shall investigate two different methods for finding clusters in our
simulated maps.  The first simply flags local maxima in the (smoothed)
map.  The second uses a matched filtering algorithm similar in spirit
to the (more sophisticated) algorithm discussed in
Hobson \& McLachlan (\cite{HobMcL}) and in implementation to the
matched filter described in White \& Kochanek (\cite{WhiKoc}).

The first, and simplest, method starts with a list of local maxima.
Since our pixel scale is $\ll 5'$ we first smooth the maps to reduce the
noise level per pixel.  We choose an additional smoothing of the same size
as the beam.  Fortunately, our results are not very dependent on this choice.
If we increase the smoothing slightly we find more nearby clusters at the
expense of distant clusters.  The optimal filtering appears to be $20-40\%$
larger than the beam size, though it does depend on the noise level chosen.
The smoothing filter could also be combined with a high pass filter (for
example using the mexican hat wavelet; e.g.~Cayon et al.~\cite{CSBMVTSDA},
Herranz et al.~\cite{HSHBDML}) without changing our results.
While an optimization of the filter would be an interesting exercise,
the differences are slight and we shall neglect this extra degree of
freedom here.  Once we have a list of candidate clusters, we keep all
whose peak value in the map lies above a given threshold, discussed below.
We also keep track of which peaks match halos in our 3D catalogues.
When matching peaks and halos we require that the center of the halo lie
within $2'$ (4 pixels) of the peak maximum.

The second method is a combination of a matched filter algorithm and a
`CLEAN'-like algorithm (H\"ogbom \cite{Hog}; Clark \cite{Cla}).
The key assumption is that clusters will look like point sources at the
{\sl Planck\/} resolution.
Assuming that the clusters of interest are unresolved, we model clusters as
Gaussian with the same width as the beam.  We do not scan over a range of
filter sizes as in Herranz et al.~(\cite{HSHBDML}).
Given a map we evaluate the best fitting amplitude of a Gaussian for each
pixel.  The fit is done over $3'$ around the central pixel to avoid
overweighting the outskirts of the cluster\footnote{Note that since we
are fitting to a profile which is already smoothed by the beam, we are
however including emission from much larger than this radius.}.
Our results are not very sensitive to this choice.
The most significant detection is added to our peak list, and then the best
fit Gaussian is subtracted from the map.  This process is iterated until no
further significant detections ($>3\sigma$) are found.
One advantage of this method is that it can easily deal with edges and
non-uniform noise while retaining the full resolution of the map (i.e.~we
avoid any additional smoothing).  A disadvantage is that we implicitly assume
the clusters are unresolved (i.e.~beam shaped).
As above, when matching peaks and halos we require that the center of the
halo lie within $2'$ (4 pixels) of the peak center.

\begin{figure}
\begin{center}
\resizebox{3.5in}{!}{\includegraphics{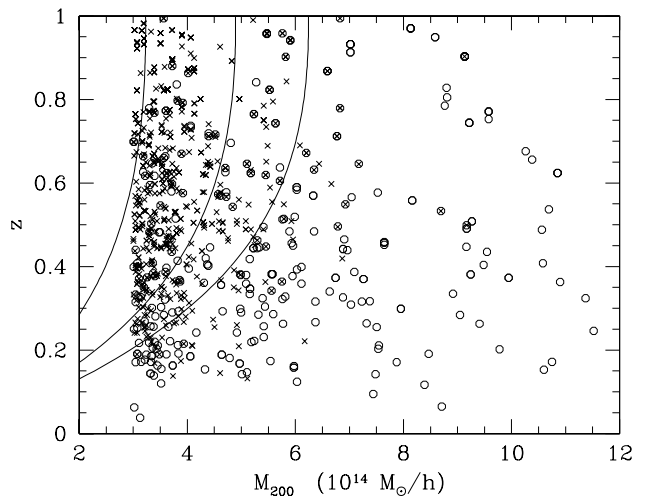}}
\end{center}
\caption{The clusters found and missed by matching peaks in maps with
$20\,\mu$K of noise.  We have kept all peaks above 50\% reliability for
halos more massive than $3\times 10^{14}\,h^{-1}M_\odot$.  Crosses indicate
clusters in the field which did not match any peak above the threshold,
open circles the clusters which lay within $2'$ of a selected peak.
Crosses inside circles indicate clusters which were found in one orientation
but missed in another (corresponding to different lines-of-sight with
differing large-scale structure in projection etc).
The solid lines show (left to right) lines of constant SZE flux
(200mJy, 400mJy and 600mJy) in the idealized model described in the text.}
\label{fig:mz}
\end{figure}

\begin{figure}
\begin{center}
\resizebox{3.5in}{!}{\includegraphics{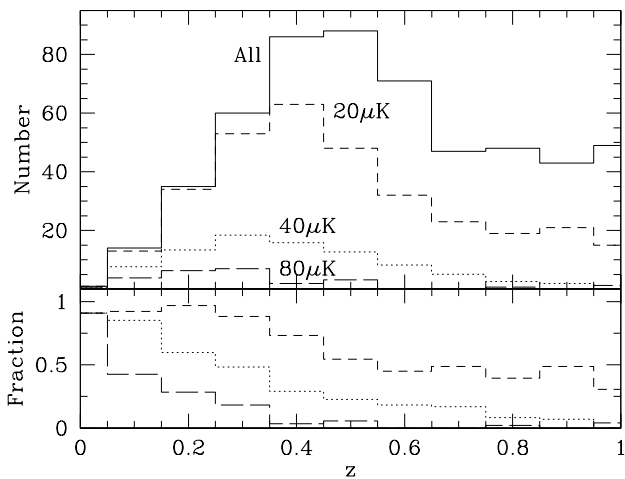}}
\end{center}
\caption{The redshift distribution of clusters above
$5\times 10^{14}\,h^{-1}M_\odot$.  (Top)  The solid histogram shows all of
the clusters in the 10 fields, which agrees quite well with the predictions
of Press-Schechter theory (not shown) once masses are corrected to $M_{200}$.
The dashed and dotted lines show those clusters which lay within $2'$ of a
peak above the threshold for the indicated level of noise in the map.
(Bottom) The fraction of clusters found.}
\label{fig:histz}
\end{figure}

\section{Results} \label{sec:results}

We present our results in terms of the completeness and efficiency
(or reliability) of the method in finding clusters above a mass threshold.
Completeness is the ratio of the number of clusters we found using the mock
SZE observation to the total number of massive clusters in the field of view.
Out of the total number of cluster candidates that we identify in our SZE
maps, only some of them will actually be clusters with a mass above the
threshold of interest.
We allow more than one cluster to match a peak, under the assumption that
any confirmation procedure would likely detect both clusters and thus we
would have `found' both.
On the other hand, we make no correction for edge effects so our
completeness will typically be underestimated by a few percent.
Efficiency (or reliability) will measure the ratio of clusters found to
the total number of candidates, and is a measure of the amount of
contamination suffered when using the SZE technique.

Often completeness and efficiency are phrased not in terms of finding
clusters above a fixed mass but rather in terms of finding clusters above
a certain flux threshold.  Since there is scatter in the relation between
mass and flux the conversion between these definitions is non-trivial.
Even a sample which is 100\% complete and efficient in flux selection will
be incomplete above a given mass threshold if the scatter is non-zero.
In terms of the physical properties of the clusters that {\sl Planck\/}
finds the selection in terms of mass is most useful, and that is why we
focus on it here.  It is possible, indeed likely, that for characterizing
the sample and selection effects at a later stage a flux limit may be
more appropriate.

Despite the added complexity we find that our matched filter method performs
worse than the simple peak finding on smoothed maps, missing more of the
high mass clusters.  This suggests that at $5'$ our assumption that all of
the SZE signal is unresolved, or that the peaks are beam shaped, is not true.
To improve upon the method would require us to perform an expensive search
through non-circular, arbitrary sized shapes.
For this reason we shall concentrate here primarily on the results from the
peak finding algorithm.
We find that for all levels of noise and smoothing in the maps our
efficiency/reliability drops rapidly as we lower the candidate threshold.
We shall adjust our threshold for each map so that our reliability is 50\%
(and we have searched at least 10 peaks).
The precise value this takes will depend on the noise, smoothing and
the mass threshold for halos we are considering.  For a cut at
$5\times 10^{14}\,h^{-1}M_\odot$ these thresholds are approximately
40, 80 and $160\,\mu$K, with a slight variation from map to map.
It is unlikely that a lower threshold would be chosen, so this is the
most optimistic assumption in terms of finding clusters.

We show in Fig.~\ref{fig:mz} the distribution of halos which are found
in our lowest noise fields.  We see immediately that {\sl Planck\/} finds
all of the highest mass clusters in the field out to $z\simeq 0.8$, but
only about half of the clusters at lower masses even with this optimistic
noise level.  Based on this figure we shall set our mass threshold for
candidates at $5\times 10^{14}\,h^{-1}M_\odot$.  This is somewhat larger
than thresholds which have been quoted in earlier papers, but based on
Fig.~\ref{fig:mz} we expect a small fraction of the clusters less massive
than this will be detected by {\sl Planck} using this peak finder.
We also show on Fig.~\ref{fig:mz} some lines of constant SZE `flux',
assuming infinite resolution.
These lines are computed following Kay, Liddle \& Thomas (\cite{KayLidTho})
Eq.~(14), which gives the integral of the SZE over the virialized region
of the (isothermal) cluster as
\begin{eqnarray}
  S_\nu(353\,{\rm GHz}) &\simeq& 75\,{\rm mJy}
  \ \left({M_{200}\over 10^{14}\,h^{-1}M_\odot}\right)^{5/3} \\ \nonumber
  &&
  \times \left({D_A\over 500\,h^{-1}{\rm Mpc}}\right)^{-2} (1+z)
\end{eqnarray}
where $D_A$ is the angular diameter distance.
Beware that these lines do not include the
effects of the finite resolution of {\sl Planck\/}.  They are included to
show how closely the selection corresponds to an intrinsic flux cut.

There are some clusters which appear in Fig.~\ref{fig:mz} as both `found'
and `missed'.  For these clusters the detection depends upon the orientation
of the line-of-sight, and we have simulated several orientations.
Due to a combination of line-of-sight projection and asphericity effects the
cluster is easier to find in some orientations than others.
This kind of effect is absent in analytic models based on spherical clusters
placed randomly in background fields.

Finally we show the redshift projection in Fig.~\ref{fig:histz}.
Above $5\times 10^{14}\,h^{-1}M_\odot$ the completeness is relatively
good for `local' clusters, but drops below 50\% beyond $z\sim 0.5$.
Since the volume is a rapidly rising function of redshift many of the
{\sl Planck\/} clusters will lie beyond $z=0.5$, they will not however be a
very complete sample.

Our investigations suggest that the noise level is quite important in the
{\sl Planck\/} cluster yield.  We find most of the massive clusters in the
deepest, cleanest parts of the sky (our $20\,\mu$K example).
This degrades somewhat by $40\,\mu$K.
By $80\,\mu$K we are finding very few of even the most massive clusters.
Our estimate of the unresolved point source contribution, from the SCUBA
counts, suggests that the latter situation is the most likely over much
of the sky.

\section{Working at lower frequency} \label{sec:lowfreq}

If our estimates of the signal strength and foreground contamination are
correct then (at least for simple peak finding) {\sl Planck\/} will be
`noise' limited over most of the sky, with the noise dominated by unresolved
IR point sources or residual dust emission.
This suggests bringing in information from lower frequencies, where these
foreground contributions are less (see also the discussion in
Herranz et al.~\cite{HSHBDML}).
The lower frequency foregrounds would then have to be controlled
by the channels at $100\,$GHz and below (which emphasizes the importance of
the wide frequency coverage of {\sl Planck}) or perhaps reduced by high-pass
filtering of the maps to remove diffuse emission.

For both the $143$ and $217\,$GHz channels the unresolved IR point source
contributions are comparable to the instrument noise in those channels
while the lower frequency foregrounds are expected to be below the noise.
Thus if we difference the $143\,$GHz and $217\,$GHz channels to isolate CMB
{}from SZ, and we assume that the lower frequency channels have been used
to remove the residual low frequency emission, we will be dominated by the
noise (and foregrounds) in the $217\,$GHz channel: we will use $15\,\mu$K per
$7'$ pixel.  Our SZ signal will now come from $143\,$GHz which gives
thermodynamic temperature fluctuations a factor $\sim 2$ lower at the same
$y$ value.
Our procedure is exactly as outlined above, except that we increase the
matching radius to $3'$ (6 pixels) since our smoothing length has
correspondingly increased.  This relaxed matching criterion, which allows
40\% more sky area, should be borne in mind when comparing the different
frequencies.

\begin{figure}
\begin{center}
\resizebox{3.5in}{!}{\includegraphics{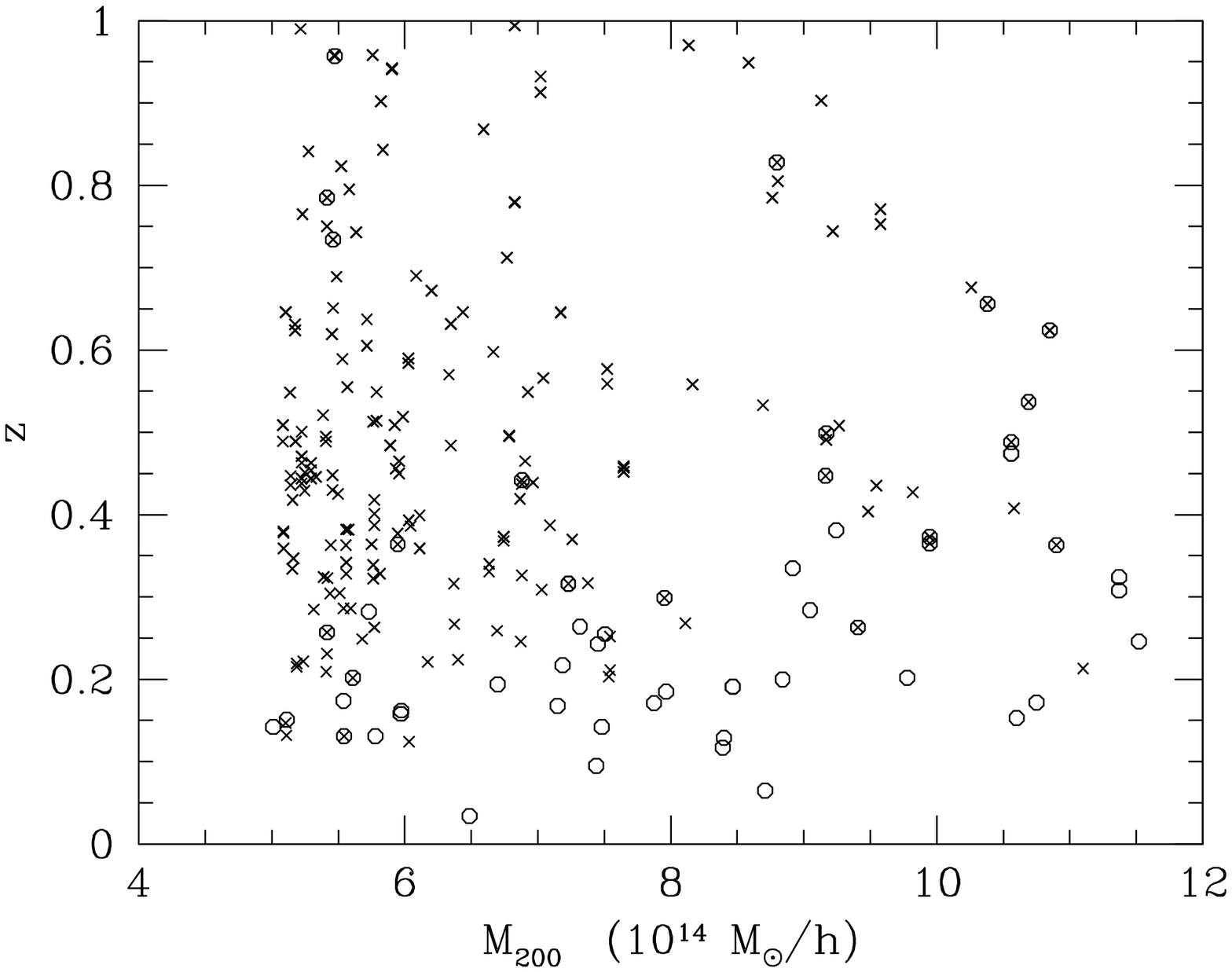}}
\end{center}
\caption{The clusters found and missed by matching peaks in our low-frequency
maps with $15\,\mu$K of noise.  We have kept all peaks above 50\% reliability 
for halos more massive than $5\times 10^{14}\,h^{-1}M_\odot$.  Crosses indicate
clusters in the field which did not match any peak above the threshold,
open circles the clusters which lay within $3'$ of a selected peak.
Crosses inside circles indicate clusters which were found in one orientation
but missed in another (corresponding to different lines-of-sight with
differing large-scale structure in projection etc).}
\label{fig:lomz}
\end{figure}

At $7'$ the correspondence between peaks in the temperature map and massive
clusters is becoming loose.  The large smoothing means that the cluster signal
is significantly diluted and clusters in overdense regions have a higher
chance of being included in the sample than relatively isolated clusters.
Our matched filter now does as well as simply smoothing the maps and looking
for local maxima, suggesting that the peaks are now well approximated by the
shape of the beam, but it still does not perform significantly better so we
shall stick to the simpler method for comparison with \S\ref{sec:results}.
Fig.~\ref{fig:lomz} illustrates the situation in the same format as
Fig.~\ref{fig:mz}.
The completeness is intermediate between the $5'$ cases with $40\,\mu$K
and $80\,\mu$K noise across the entire range of redshifts and we do not
show it explicitly.

This suggests that a {\sl Planck\/} sample obtained from the lower
frequencies is indeed an option should the situation at higher frequencies
be closer to (or higher than) the $80\,\mu$K figure than the $40\,\mu$K in
the patch of sky under consideration.
Viewed another way, the overlap of two catalogues produced at these two
frequencies would provide an important cross check.
We defer consideration of optimal combinations of the channels to a future
paper.

\section{Follow up} \label{sec:followup}

\begin{figure}
\begin{center}
\resizebox{3.5in}{!}{\includegraphics{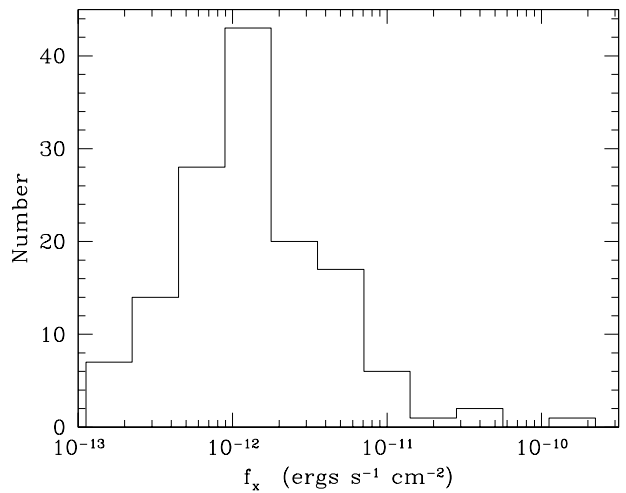}}
\end{center}
\caption{The expected X-ray flux of clusters detected as peaks in the
$353\,$GHz channel with $40\,\mu$K of noise per $5'$ beam in our 10
maps of 100 sq.~deg.~each.}
\label{fig:fxhist}
\end{figure}

The majority of the {\sl Planck\/} cluster sample will be unresolved, so the
exploitation of the data for cluster science rests on their combination and
correlation with independent data sets.  For this reason, and more generally,
it is of some interest to ask what are the optical and X-ray properties of
the cluster sample we have identified here.  These properties will determine
the optimal follow-up procedures and set the scope for any pre-launch work
which could aid in cluster finding (see also Diego et al.~\cite{DMSBS}).

We will use simple models to make a rough estimate of the X-ray, optical and
lensing properties of {\sl Planck\/} clusters
(see Table \ref{tab:clusterprop}).
Firstly, a cluster of $10^{15}\,h^{-1}M_\odot$ has a (1D) velocity dispersion
of $1100$km/s at $z=0$, scaling as $(E(z)M_{200})^{1/3}$ where
$E(z)=H(z)/H_0$ is the evolution parameter or dimensionless Hubble
parameter.
For the X-ray properties we make use of measured scaling relations between
mass, temperature and luminosity.
Given the mass and redshift of a cluster we compute the cluster temperature
{}from (Finoguenov, Reiprich \& B\"{o}hringer \cite{FRB})
\begin{equation}
  {T\over 1{\rm keV}} = \left(
    {M_{500}\over 2\times 10^{13}\,h^{-1}M_\odot} \right)^{2/3}\ E^{2/3}(z)
\end{equation}
where the redshift evolution has been taken to follow self-similar collapse.
The conversion from temperature to luminosity is still somewhat uncertain,
depending on assumptions made about cooling flows and other corrections.
For example, Allen \& Fabian (\cite{AllFab}; Model C) quote a conversion
between temperature and (bolometric) luminosity as
\begin{equation}
  L = 5\times 10^{44}\left( {T\over 6\,{\rm keV}} \right)^{2.33}
  \ h^{-2} {\rm ergs}\,{\rm s}^{-1} \qquad ,
\end{equation}
For comparison, a slightly different conversion is given by
Markevitch (\cite{Mar})
\begin{equation}
  L = 3\times 10^{44}\left( {T\over 6\,{\rm keV}} \right)^{2.64}
  \ h^{-2} {\rm ergs}\,{\rm s}^{-1} \qquad .
\end{equation}
We shall use the Markevitch relation, with its more conservative assumptions,
and assume that the $L-T$ relation does not evolve with redshift,
but the difference between these two relations should be taken as an
indication of the uncertainty.
We apply the bolometric and K-corrections from Romer et al.~(\cite{RVLM})
and compute the flux from the X-ray luminosity through
\begin{equation}
  f = {L\over 4\pi D_L^2}
\end{equation}
where $D_L$ is the luminosity distance to the cluster redshift.
The corrections work at the few percent level, compared to the exact
calculations using \textsc{\small XSPEC} (Arnaud \cite{XSPEC})
which is more than adequate for our purposes
(A.K.~Romer, private communication).
Throughout we shall quote fluxes in the $0.5-2.0$keV band.
Using these scalings the distribution of fluxes is as shown in
Fig.~\ref{fig:fxhist}, peaking near
$10^{-12}\,{\rm ergs}\,{\rm s}^{-1}\,{\rm cm}^{-2}$
with a tail to $10^{-13}$ and a bright end as high as $10^{-10}$.
Many of these clusters will be detectable in reasonable (several ksec)
integrations with existing X-ray telescopes if they are still operating.
Of the existing X-ray surveys,
those based on the RASS
(Tr\"{u}mper et al.~\cite{RASS91}, Voges et al.~\cite{RASS99};
e.g.~Reflex: Bohringer et al.~\cite{Reflex};
     MACS: Ebeling, Edge \& Henry \cite{MACS};
     NEP:  Henry et al.~\cite{NEP};
     BCS: Ebeling et al.~\cite{BCS})
have the combination of sensitivity and sky coverage to provide the best
pre-launch catalogues for {\sl Planck}.  In the fields where {\sl XMM\/}
and {\sl Planck\/} overlap there should be strong X-ray detections in the
XCS (Romer et al.~\cite{RVLM}).
There will additionally be many clusters in the deep optical surveys which
will be completed by the time {\sl Planck\/} launches (see below).

\begin{figure}
\begin{center}
\resizebox{3.5in}{!}{\includegraphics{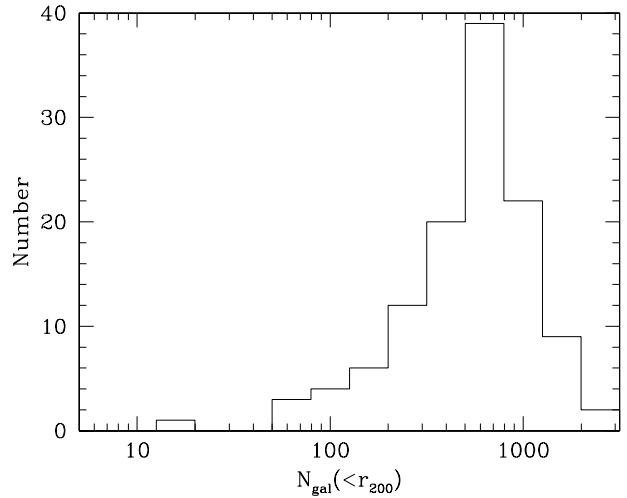}}
\end{center}
\caption{The number of galaxies brighter than $R=25$ within the virial
radius of our {\sl Planck\/} cluster sample.  The number within the core
region, where the contrast against the background is higher, would be
about 10\% of $N_{\rm gal}(<r_{200})$.  Again the clusters are those in the
$353\,$GHz channel with $40\,\mu$K of noise per $5'$ beam and 10 maps of
100 sq.~deg.~each.}
\label{fig:nhist}
\end{figure}

Optical and near-IR emission is still the least expensive way of measuring
cluster redshifts.  Without redshift information, the distance to the cluster
is essentially unknown and so the physical interpretation of the cluster sample
depends on our ability to detect clusters in optical light and determine
their redshifts.

\begin{figure}
\begin{center}
\resizebox{3.5in}{!}{\includegraphics{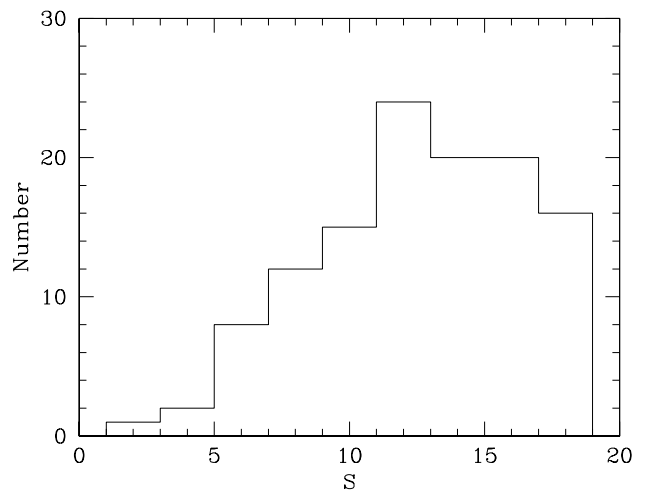}}
\end{center}
\caption{The weak lensing signal-to-noise, assuming the parameters described
in the text, of clusters detected as peaks in the $353\,$GHz channel with
$40\,\mu$K of noise per $5'$ beam and 10 maps of 100 sq.~deg.~each.}
\label{fig:shist}
\end{figure}


In order to determine how many clusters we could find or study from
optical imaging it is useful to have an estimate of the cluster richness.
For our purposes a rough estimate suffices, so we model the (R-band)
luminosity function of cluster galaxies after that of Coma, for which a
Schechter function with $M_\ast=-20.9$ and $\alpha=-1.2$ is a reasonable
fit (Beijersbergen et al.~\cite{ComaLF}).
The normalization should scale roughly with the mass, $\phi_\ast\propto M$,
but the overall value is uncertain.  We will assume 20 galaxies within the
virial radius brighter than $L_\ast$ for a $10^{15}\,h^{-1}M_\odot$ halo.
The number in the core region, where the contrast against the background is
the highest, will obviously be smaller.
If the galaxies follow the mass, approximately 20\% of the galaxies lie within
the break radius ($0.2r_{200}$) and 8\% within the core radius ($0.1r_{200}$)
of a rich cluster.
Assuming pure luminosity evolution, $L_\ast$ scales as $10^{0.4z}$
(van Dokkum et al.~\cite{Dokkum}).
For simplicity we use the K-corrections appropriate to elliptical galaxies
for the Sloan $r'$ band from Fukugita, Shimasaku \& Ichikawa (\cite{FSI}).

We see from Fig.~\ref{fig:nhist} and Table \ref{tab:clusterprop} that we
will find a significant number of very massive, high-$z$ clusters which
should contain many galaxies within the virial radius brighter than $R=25$
($5\sigma$ photometry to $R=25$ takes approximately 1 hour on a 4m class
telescope).
Even taking into account that we will have to work at a small fraction of
the virial radius, the {\sl Planck\/} sample will be ideal for studying the
galaxies in the most massive and distant clusters.
Since the {\sl Planck\/} sample is biased towards the most massive clusters,
which should contain a red-sequence, this also suggests that deep 2-band
photometry would be an efficient way to verify cluster candidates and obtain
an approximate redshift (Gladders \& Yee \cite{GlaYee}) for those which have
not already been detected by SDSS (Bartelmann \cite{MB}) or other galaxy
surveys.
Once the candidates are confirmed as clusters, with an approximate redshift,
it will be possible to target a sub-sample for spectroscopy.  Above $R=22$,
which is achievable on a 10m telescope with a moderate (30min) integration,
the number of galaxies is approximately $1/3$ of the number shown in
Fig.~\ref{fig:nhist}.

The gravitationally lensing subsample of the {\sl Planck\/} clusters has
been investigated by Bartelmann (\cite{MB}), who showed that many of the
clusters {\sl Planck\/} will detect should also give measurable weak lensing
signatures.  Following Bartelmann we have computed the $S$-statistic, giving
the signal to noise for 30 galaxies per arcmin${}^2$ at $z=1.5$, for the weak
lensing signal on $1'$.  This is optimistic for current observations, but may
be attainable by future facilities.  Figure \ref{fig:shist} shows that almost
the entire {\sl Planck\/} cluster sample will be excellent targets for weak
lensing surveys, having extremely high signal to noise.
Though line-of-sight projection can be a significant contaminant for weak
lensing studies of the general cluster population
(Metzler, White \& Loken \cite{MetWhiLok};
 White, van Waerbeke \& Mackey \cite{WhivWaMac})
it is less of a problem for the {\sl Planck\/} sample which consists primarily
of the most massive systems.
Our conclusions are thus in line with, though slightly stronger than, those
of Bartelmann who suggested that {\sl Planck\/} clusters would form an
excellent sample for weak lensing follow-up.  It also follows from this that
many of the {\sl Planck\/} clusters will produce measurable lensing features
in the CMB itself, although the resolution of {\sl Planck\/} is not well
matched to mapping the majority of these clusters.

\begin{table}
\begin{center}
\begin{tabular}{ccccccc}
$M_{200}$ & $z$ & $L_X$ & $f_X$ & $T_X$ &  $S$  & $N_{\rm gal}$ \\
\hline
   3      & 0.2 &  0.7  &  1.3  &  5.2  &  7    & 350 \\
   3      & 0.4 &  0.9  &  0.3  &  5.6  & 11    & 190 \\
   3      & 0.6 &  1.0  &  0.1  &  6.1  & 11    &  90 \\
   3      & 0.8 &  1.3  &  0.0  &  6.6  & 11    &  45 \\
   3      & 1.0 &  1.5  &  0.0  &  7.1  &  8    &  20 \\
  10      & 0.2 &  2.9  &  5.0  &  11   & 10    &1100 \\
  10      & 0.4 &  3.5  &  1.3  &  12   & 15    & 600 \\
  10      & 0.6 &  4.3  &  0.6  &  13   & 17    & 300 \\
  10      & 0.8 &  5.2  &  0.4  &  14   & 16    & 150 \\
  10      & 1.0 &  6.4  &  0.2  &  15   & 12    &  60
\end{tabular}
\end{center}
\caption{Cluster properties, computed with our simple scaling laws.
The mass is quoted in $10^{14}\,h^{-1}M_\odot$, luminosity in
$10^{44}\,h^{-2}\,{\rm ergs}\,{\rm s}^{-1}$,
flux in $10^{-12}\,{\rm ergs}\,{\rm s}^{-1}\,{\rm cm}^{-2}$,
temperature in keV, lensing as a SNR and number of galaxies within
the virial radius brighter than $R=25$.}
\label{tab:clusterprop}
\end{table}

We show the cluster properties from our simple model in
Table \ref{tab:clusterprop} for a range of redshifts.
Because we are holding the mass fixed, the X-ray temperature (and thus
luminosity) are increasing functions of redshift.  The lensing signal
peaks half way between $z=0$ and $z=1.5$, around $z\simeq 0.6$.

\section{Cosmology dependence} \label{sec:model}

We have assumed a particular cosmological and foreground model for our
study, and it is important to ask how sensitive our results are to this
assumption.  Our neglect of dust emission is clearly optimistic.
Our `foreground noise' is matched to the SCUBA counts at $353\,$GHz, but
these counts are uncertain at the 50\% level, leading to a similar
uncertainty in our noise level.  We have assumed a fairly steep frequency
dependence in extrapolating to lower frequencies, and this may not be true
of all the sources.  Conversely, the brightest IR sources may also be
bright in some other waveband (e.g.~radio), which may provide additional
leverage in removing them.
On the cosmological side the $\Lambda$CDM paradigm seems to be fairly secure,
and the values of the matter density and Hubble constant we have chosen are
quite standard.
However all of the calculations presented above had the SZ signal normalized
to match the CBI and BIMA results.  If the CBI signal is not dominated by
thermal SZ then the SZ signal from clusters {\it may\/} be lower than we
have assumed.
This scenario would suggest using a lower matter power spectrum normalization
($\sigma_8$) than we have used throughout.

To this end we have run another simulation, identical to that described in
\S\ref{sec:hifreq}, except that $\sigma_8=0.8$.
As expected the lower $\sigma_8$ reduces the number of the most massive and
high redshift clusters.
Our redshift distributions are then shifted slightly to lower $z$.
The number of massive clusters which {\sl Planck\/} can see is reduced
compared to the number of lower mass SZE sources, which slightly increases
the confusion.
Neither of these two effects alters our conclusions, so the main impact is
that there are fewer clusters overall.
Simply rescaling the total number of clusters doesn't have a large impact
on our conclusions, it only makes our statistics more susceptible to
Poisson fluctuations from the relatively small simulation volume.
Thus running the simulation with $\sigma_8=1$ can be regarded as a simple
way to increase the statistics on rare objects while keeping the box size
small and hence the mass and force resolution high.

We have used a cluster mass-temperature normalization close to that predicted
by the hydrodynamic simulations, chosen to match the level of power seen by
CBI\footnote{Most of the increase in power in our simulations compared to
the hydrodynamic simulations comes from our high assumed $\sigma_8$ since
at fixed $M-T$ normalization $C_\ell\propto\sigma_8^{14/(3+n)}$ where $n$
is the effective spectral index.}.  However, X-ray observations suggest that
clusters at a fixed mass may be hotter than most simulations predict
(see e.g.~Table 1 of Muanwong et al.~\cite{MTKP}
or Fig.~2 of Huterer \& White \cite{HutWhi}).
If this holds across the whole cluster, {\it and\/} the temperature estimated
by X-ray observations is close to a mass weighted temperature, this would
imply an increase in the SZ signal per cluster.
An increase, at the 20-50\% level, in the SZ signal per cluster is possible.
A simple rescaling of the noise, which is anyway uncertain at this level,
would mimic any such change.

Thus we expect that our simulations, while not perfect, should provide a
fair guide to the expected size of the SZ signal based on our current
knowledge.  If anything the indications are that our simulations may predict
too many clusters, especially at high redshift, and that each cluster may
produce slightly too little SZ signal.  At present it is not unreasonable
to assign a factor of 2 uncertainty to the signal-to-noise in our maps.

\section{Conclusions} \label{sec:conclusions}

The SZE offers a new and potentially very powerful method for finding high
redshift clusters of galaxies, and the {\sl Planck\/} mission will be
unique in producing all-sky maps at the relevant frequencies with high
angular resolution and sensitivity.

We have presented a preliminary investigation of the cluster sample
which {\sl Planck\/} should provide.  We used mock SZ maps drawn from a large
volume, high resolution N-body simulation.  These maps capture much of the
physics behind the SZ effect, with the sources (groups and clusters) situated
correctly in their cosmological context.  To these maps we add `noise'
arising from the detectors and from incomplete foreground subtraction.
Our foreground modeling has been highly simplistic and idealized, both because
cluster finding is a local process and because we wish to decouple that
degree of uncertainty from the main focus of this work.
Uncertainty in cluster physics and our foreground model suggests that our
signal-to-noise may be uncertain at (up to) the factor of 2 level.
For this reason we have simulated a range of `effective' noise amplitudes
at fixed signal.
Improving our normalization of the sources will require a moderate sample of
clusters whose integrated SZ signal and X-ray temperature or velocity
dispersion is known, for comparison with the simulations used here.
Our high frequency foreground uncertainty will be improved by better
measurements of the dust emission and the bright end of the source luminosity
function and the frequency dependence of the sources.
The sources of interest for {\sl Planck\/} are the rarer sources brighter
than a few tens of mJy.

We have used combinations of only 2 frequency maps, plus vetoing regions where
the $545$ and $857\,$GHz maps show strong dust/sources, to disentangle the SZ
signal from the CMB and astrophysical foregrounds.
This is perhaps the simplest method, and the easiest to understand.
It remains to be seen how it compares with more complex algorithms and
more realistic treatments of foregrounds.

We have found cluster candidates in our simulated difference maps either by
flagging local maxima or using a point-source optimized matched filter
algorithm.
We find the former is at least as successful as the latter, suggesting that
even at $5'$ a significant fraction of the detections will not be beam
shaped and would be missed by algorithms optimized to find point sources.
For our lowest noise levels we show that {\sl Planck\/} detects almost all
of the nearby, massive clusters and a significant fraction of the massive
clusters out to $z\sim 1$.  The completeness in our 10 fields ranges from
$40-70\%$ for all clusters above $5\times 10^{14}\,h^{-1}M_\odot$.
At intermediate noise levels the range is $10-30\%$ while at the highest
noise level we consider (our fiducial model), our algorithm recovers only
a small fraction of even the richest clusters ($2-10\%$).
We expect the foreground contamination to be less at lower frequencies, and
even with the corresponding decrease in angular resolution cluster finding
is better done with $217\,$GHz and below if the signal-to-noise ratio in a
given region of sky is as low as in our fiducial model.
The combination of candidates from both the high and low frequency methods,
perhaps used in different parts of the sky, would provide an interesting cross
check on foreground contamination which we shall defer to future work.
The uncertainties in the amplitude of the power spectrum, the signal and noise
levels, the sky distribution of the foregrounds and the optimal methods for
identifying clusters make predictions of the number of clusters that will
be found by {\sl Planck\/} highly uncertain.  Nominally {\sl Planck\/} should
detect from several thousand to more than ten thousand rich clusters.
Even very pessimistic assumptions give the return at several hundred clusters,
suggesting that the {\sl Planck\/} catalogue will be a fertile ground for
future investigations.

Using simple minded scaling arguments we have estimated the X-ray, optical
and weak lensing properties of the {\sl Planck\/} cluster sample.  We find
that most of the sample will be strong weak lensing sources, and will
contain a large number of galaxies.  The X-ray emission from many sources
will be detectable with relatively deep integrations with existing facilities
should they still be operational.
However, given the high-$z$ and all-sky nature of the expected cluster yield,
pre-launch samples will most likely be constructed with dedicated SZ
instruments, optical or weak lensing surveys.

An all-sky sample of massive clusters with a well understood selection
function would be a goldmine for cosmology.  {\sl Planck\/} should be
the first mission capable of producing such a catalogue.  However the
$5'$ resolution of {\sl Planck\/}, along with noise from astrophysical
and detector sources, make understanding of the selection function highly
non-trivial.  Better algorithms for extracting clusters from the
multi-frequency maps need to be developed and more realistic simulations
performed before precision cosmology can be extracted from future
{\sl Planck\/} SZE cluster observations.  To aid in this endeavor the
maps, cluster catalogues and derived quantities from this work are available
to the public at http://mwhite.berkeley.edu/.  It is to be hoped that these
simulations will be replaced by deep, high resolution, wide field SZE survey
data as it becomes available.

\bigskip
I would like to thank A.K.~Romer for helpful discussions on X-ray fluxes and
K-corrections, H.~Mo and D.~Eisenstein for discussions on optical
K-corrections, M.~Davis for discussions on spectroscopic follow-up
and C.~Kochanek for numerous discussions on cluster related issues.
I thank A.~Amblard for help with the {\sl IRAS\/} data and
J.~Diego, M.~Hobson, C.~Lawrence and D.~Scott for helpful comments on an
earlier draft of this work.
The simulations used here were performed on the IBM-SP2 at the National
Energy Research Scientific Computing Center.
This research was supported by the NSF and NASA.

\end{document}